\documentstyle[12pt, aasms4]{article}
\begin{document}

\vspace*{2.0in}
\title{Gas and dust in NGC 7469: sub-mm imaging and CO J=3--2}

\author{Padeli P. Papadopoulos\altaffilmark{1}}

\and

\author{Michael L. Allen \altaffilmark{2}}

\altaffiltext{1}{Sterrewacht Leiden, P. O. Box 9513, 2300 RA Leiden, 
The Netherlands}
\altaffiltext{2}{Department of Astronomy, University of Toronto, 60
St. George street, Toronto,\\ \hspace*{0.5cm} ON M5S-3H8, Canada}

\begin{abstract}

We present sensitive  sub-mm imaging of the Seyfert  1 galaxy NGC 7469
  at 850  $\mu $m and 450  $\mu $m with the  Submillimetre Common User
  Bolometer Array (SCUBA) on  the James Clerk Maxwell Telescope (JCMT)
  and $ ^{12}$CO J=3--2  line observations of its central starbursting
  region.  The global dust spectrum,  as constrained by the new set of
  sub-mm data and  available 1.30~mm and IRAS 100~$\mu  $m, 60 $\mu $m
  data reveals  a dominant warm  dust component with a  temperature of
  $\rm T_{\rm d}  \sim 35$ K and a  global molecular gas-to-dust ratio
  $\rm M(H_2)/M_{\rm d}\sim 600$.   Including the atomic gas component
  yields a  total gas-to-dust ratio  of $\sim 830$.  Such  high values
  are typical  for IR-bright  spirals and in  order to  reconcile them
  with the  significantly lower ratio  of $\sim 100$ obtained  for the
  Milky Way  a cold dust reservoir, inconspicuous  at FIR wavelengths,
  is usually  postulated.  However, while  there is good  evidence for
  the presence of cold gas/dust in NGC~7469 beyond its central region,
  our 450 $\mu $m map  and available interferometric $ ^{12}$CO J=1--0
  maps show the bright sub-mm/CO  emission confined in the inner $\sim
  2.5$~kpc,  where a  high $  ^{12}$CO (J=3--2)/(J=1--0)  ratio ($\sim
  0.85-1.0$) is  measured.  This is  consistent with molecular  gas at
  $\rm T_{\rm kin}\ga 30$K, suggesting that the bulk of the ISM in the
  starburst   center   of   NGC~7469   is  warm.    Nevertheless   the
  corresponding  total gas-to-dust  ratio there  remains high,  of the
  order of  $\sim 500$.  We  argue that, rather than  unaccounted cold
  dust mass, this high ratio  suggests an overestimate of $\rm M(H_2)$
  from its associated $ ^{12}$CO J=1--0 line luminosity by a factor of
  $\sim  5$  when a  Milky  Way value  for  this  conversion is  used.
  Finally the diffuse  cold gas and dust that is  the likely source of
  the  observed faint  extended  450  $\mu $m  and  $ ^{12}$CO  J=1--0
  emission has an estimated  total gas-to-dust ratio of $\sim 50-160$,
  closer to the Galactic value.

\end{abstract}


\section{Introduction}

The role of  molecular gas as the fuel of  both starburst activity and
an Active Galactic Nucleus (AGN)  is now well established.  In Seyfert
galaxies such gas is found on scales ranging from the inner $\sim 1-2$
kpc ``feeding''  an intense circumnuclear starburst down  to $\rm L\la
$100~pc from the AGN, where in the form of a geometrically thick torus
obscures the active nucleus along certain viewing angles thus creating
the difference between type 2 and type 1 Seyferts (Miller \& Antonucci
1983; Krolik 1990 and references therein).

Estimates  of  molecular  gas   mass  in  other  galaxies  from  their
 velocity-integrated $ ^{12}$CO J=1--0 luminosity $\rm L_{\rm CO}$ (in
 K km s$ ^{-1}$ pc$ ^2$)  are based on the so-called conversion factor
 $\rm  X_{\rm  CO}=M(H_2)/L_{\rm  CO}\approx  5\ M_{\odot  }  (K\  km\
 s^{-1}\  pc^2)^{-1}$ (e.g. Solomon  \& Barrett  1991) whose  value is
 derived from studies of molecular gas in the Galactic disk.  Numerous
 studies quantify  the effects of  the various physical  conditions on
 this   factor  (e.g.    Bryant  \&   Scoville  1996;   Israel  1997).
 Particularly  in the starburst  nuclei of  very luminous  IR galaxies
 ($\rm  L_{\rm  FIR} >  10^{11}\  L_{\odot  }$)  these conditions  are
 significantly  different than  in the  disk environment  of quiescent
 spirals like  the Milky  Way.  More specifically  in the  inner $\sim
 1-2$ kpc,  where a starburst  usually occurs, the  gas differentiates
 into  two distinct phases  (e.g.  Aalto  et al.   1995) with  a warm,
 diffuse  and possibly  non  self-gravitating phase  dominating the  $
 ^{12}$CO  emission.  This  results in  a significant  overestimate of
 $\rm H_2$  mass from $\rm L_{\rm  CO}$ when a Galactic  value of $\rm
 X_{\rm CO}$ is used (Solomon et al. 1997; Downes \& Solomon~1998).

The aforementioned  conditions are expected  in the nuclear  region of
the SBa Seyfert~1  galaxy NGC~7469 (Arp 298, Mrk  1514), a luminous IR
source  with  $\rm L_{\rm  FIR}\approx  3\times  10^{11}\ L_{\odot  }$
emanating  from a  powerful starburst  deeply embedded  into  its $\rm
L\sim 1$~kpc circumnuclear region (Wilson  et al.  1991; Genzel et al.
1995).  High resolution $  ^{12}$CO J=1--0 observations revealed large
amounts of  molecular gas  within this region  (Meixner et  al.  1990;
Tacconi \&  Genzel 1996)  with mass comparable  to the  dynamical, and
there  is  significant  evidence  that  application  of  the  standard
Galactic   conversion  factor  overestimates   $\rm  M(H_2)$   in  the
circumnuclear starburst of this galaxy (Genzel et al.  1995).  In this
paper we present  sensitive sub-mm imaging of NGC  7469 at 850~$\mu $m
and  450~$\mu $m  and $  ^{12}$CO J=3--2  spectroscopy of  its central
region.   Under the  assumption of  a canonical  gas-to-dust  ratio of
$\sim 100$,  our results  confirm the overestimate  of H$_2$  gas mass
when a  Galactic value for $\rm  X_{\rm CO}$ is  used, and demonstrate
the significance of  CO spectroscopy and sub-mm imaging  in offering a
better assessment of  the molecular gas mass and  its average physical
conditions.

Throughout  this work we  adopt $\rm  H_{\circ }  = 75\  km\ sec^{-1}\
Mpc^{-1}$ and $\rm  q_{\circ } = 0.5$, which for  cz=4900 km s$ ^{-1}$
yields a  luminosity distance  of $\rm D_{\rm  L}\sim 66$ Mpc  for NGC
7469, where 1$''$ corresponds to $\sim 310$ pc.

\section{Observations}

The sub-mm observations were made on the nights of 1997 December 3 and
1998 January 15 with the Sub-mm Common User Bolometer Array (SCUBA) at
the  15-m James  Clerk Maxwell  Telescope (JCMT)\footnote{The  JCMT is
operated by  the Joint Astronomy Center  in Hilo, Hawaii  on behalf of
the  parent organizations PPARC  in the  United Kingdom,  the National
Research Council  of Canada and  the The Netherlands  Organization for
Scientific Research.}.  SCUBA is a  dual camera system cooled to $\sim
0.1$ K  allowing sensitive simultaneous observations  with two arrays.
The short-wavelength array contains  91 pixels and the long-wavelength
array 37  pixels, with  approximately the same  field of  view, namely
$\sim  2.3'$.  For  a description  of  the instrument  see Holland  et
al. (1998).

We performed  dual wavelength imaging at  450 $\mu $m and  850 $\mu $m
using the 64-point jiggle mapping mode that allows Nyquist sampling of
the field of view.   (\cite{Ho98}).  We employed the recommended rapid
beam  switching at a  frequency of  $\sim 8$  Hz and  a beam  throw of
$150''$ in azimuth.  The  pointing and focus were monitored frequently
using Uranus and CRL 618, with an expected rms pointing error of $\sim
3''$. All  maps of NGC 7469  were ``bracketed'' by  sky-dips that were
later  used to  correct for  atmospheric extinction.   Conditions were
generally excellent for sub-mm  observations with typical opacities of
$\tau _{850}\sim  0.1-0.2$ and $\tau  _{450}\sim 0.45-0.55$ throughout
our runs.

The  beam characteristics and  calibration gains  (Jy V$  ^{-1}$ beam$
 ^{-1}$) were deduced from an extensive archive of beam maps of Uranus
 and CRL 618 taken during similar periods as our observations and with
 a similar beam-throw ($120''$).   High S/N beam maps are particularly
 important for  the calibration of 450  $\mu $m images  since the beam
 shape and  the gain  can change significantly  as the  dish thermally
 relaxes.    All  jiggle   maps  were   flat-fielded,   corrected  for
 atmospheric  extinction and  edited  for bad  bolometers/integrations
 using  the  standard reduction  package  SURF  (Jenness \&  Lightfoot
 1998).  Sky-noise  was removed by  using the bolometers in  the outer
 ring of  the two  arrays (Jenness, Lightfoot  \& Holland  1998) which
 were  assumed ``looking''  only  at  sky emission.   This  is a  good
 assumption for NGC 7469 where  most of its $ ^{12}$CO J=1--0 emission
 (and presumably most of the  sub-mm emission from dust) lies within a
 $\sim 4''$ radius (Tacconi \& Genzel~1996, Tacconi et al. 2000).

The gain and beam characteristics  at 850 $\mu $m remained essentially
the same  throughout the  runs.  This is  further corroborated  by the
fact that the ``raw'' peak  and integrated intensities of the NGC 7469
images at  this wavelength agree to  within $\sim 10\%$,  which is the
expected   calibration  uncertainty.   Hence   we  co-added   all  the
noise-weighted frames after subtracting  a small gradient from one of
them, and then scale the final one with the single gain factor of $\rm
G_{850} =290$ Jy Volt$ ^{-1}$ beam$ ^{-1}$.  The deduced beam-width is
$\Theta _{\rm HPBW}^{(850)}\sim 15''$.

 The 450~$\mu $m beam maps revealed substantial gain variations ($\sim
40\%$) between  observations taken  early in the  first shift  and the
ones taken later  on, presumably due to the  thermal relaxation of the
dish. Thus before  co-adding the images of NGC~7469  we scaled them by
the appropriate  gains of  $\rm G_{450} =1170$  Jy Volt$  ^{-1}$ beam$
^{-1}$  and  $\rm G_{450}  =780$~Jy~Volt$  ^{-1}$~beam$ ^{-1}$.   This
scaling brought their peak  and integrated intensities in agreement to
within $\sim 15\%$, which is comparable to the calibration uncertainty
expected  at 450  $\mu $m.   The average  beam-width deduced  for this
wavelength is $\Theta _{\rm HPBW}^{(450)}\sim 9''$.

The $ ^{12}$CO J=3--2 line  at 345.796  GHz  was observed towards  the
nucleus of NGC 7469  with receiver B3 on 1999  August 6.  We  used the
DAS  spectrometer  with  a bandwidth of   920~MHz  ($\sim 800$  km  s$
^{-1}$), and two independent channels  centered on the line.  The chop
scheme employed   was rapid beam  switching  with a  120$''$ azimuthal
throw at the recommended frequency of 1 Hz.  The HPBW of the telescope
beam at this frequency is $\sim 14''$ with  a beam efficiency of $\eta
_{\rm mb}=0.62$ (Matthews 1999).  Typical  system temperatures were of
the order of $\rm  T_{\rm sys}\sim 600$  K.  After  subtracting linear
baselines from all    spectra their integrated  line  intensities were
found to  agree  to  within  $\sim   10\%$, the   typical  calibration
uncertainty for receiver B3. We  then co-added all of them to  produce
the final spectrum.

\section{RESULTS}

The galaxy is detected as a  bright sub-mm source at 850 $\mu $m where
it appears  marginally resolved,  as well as  at 450~$\mu $m  where it
clearly shows faint extended emission.  We convolved a high-resolution
map  of $ ^{12}$CO  J=1--0 emission  (Tacconi et  al.  2000)  from its
original resolution of $\sim 2.5''$  to the resolution of the 450 $\mu
$m map  ($\sim 9''$) and show them, together  with the 850  $\mu $m
map, in Figure 1.

\placefigure{fig1}

The  correspondence between  the bright  450  $\mu $m  and $  ^{12}$CO
J=1--0 emission is very good, suggesting that they both trace the same
ISM material.   The high-resolution interferometric  $ ^{12}$CO J=1--0
maps  (Tacconi  \&  Genzel  1996;  Tacconi et  al.~2000)  show  bright
emission arising  from a  region with a  diameter of $\rm  d\sim 8''$,
where an  intense starburst is  embedded (Wilson et al.   1991), hence
the  presence of  warm gas  and dust  is expected  there.   Indeed the
central  region of  NGC 7469,  besides  hosting an  AGN, also  harbors
several  supergiant star  formation regions  containing  numerous (few
$\times 10^4$) OB  stars each (Genzel et al.   1995), whose intense UV
light and  subsequent supernova explosions  will warm and  disrupt the
molecular clouds  present.  The warm  dust of these clouds  can easily
dominate  the global  FIR and  even the  mm/sub-mm emission  from this
galaxy.  In principle  the new sub-mm measurements allow  for a better
evaluation  of the  dust  mass and  temperature.  Thus in  Table 1  we
compile them together  with available mm and FIR data  and use them to
model the dust emission SED, shown in Figure 2.

\vspace*{1.0cm}

\centerline{EDITOR: PLACE TABLE 1 HERE}
\vspace*{0.3cm}

\placefigure{fig2}

The reported  mm/sub-mm fluxes at 1.3   mm and 850  $\mu $m  have been
corrected for  non-dust  emission from CO lines,   which can be rather
significant (see Appendix).  The fitted dust temperature and mass are:
$\rm T_{\rm d}\sim 35$ K and $\rm M_{\rm d}=2.5\times 10^{7}\ M_{\odot
}$.  It can be seen that a single dust component  fits the data rather
well,  except for  wavelengths shortward  of  60~$\mu $m where the hot
dust ($\ga 300$ K) present in the central region (Cutri  et al.  1984)
makes  a significant  contribution to the  total  flux but contains  a
negligible fraction ($\leq \rm few  \times 10^{-4}$) of the total dust
mass.  From the  $  ^{12}$CO J=1--0 map   obtained by  Tacconi et  al.
(2000), we find a total flux  of $\rm S_{\rm CO}=(275\pm 30)$~Jy~km~s$
^{-1}$, which allows an estimate the global molecular mass from

\begin{equation}
\rm M(H_2) = 2.45\times 10^3\ X_{\rm CO}\ D^2\ S_{\rm CO}
\end{equation}

\noindent
where D=66  Mpc and $\rm  X_{\rm CO}\sim 5\   M_{\odot }\ (km\ s^{-1}\
pc^2)^{-1}$.  This  yields a   mass of $\rm   M(H_2)\approx  1.5\times
10^{10}\ M_{\odot }$ and $\rm M(H_2)/M_{\rm  d}\approx 600$, a typical
value for IRAS galaxies (Young et  al. 1986; Young  et al. 1989; Stark
et al.  1986).  Including the HI gas mass $\rm M(HI)=5.7\times 10^{9}\
M_{\odot }$ (Mirabel \& Wilson 1984) yields  a total gas-to-dust ratio
of~$\sim 830$.

The presence of  warm and dense molecular   gas in the  nucleus of NGC
7469  is responsible for the   strong $ ^{12}$CO  J=3--2 line observed
towards it.  The high resolution $  ^{12}$CO J=1--0 interferometer map
allows  for  an estimate of  the  (J=3--2)/(J=1--0) line  ratio in the
inner $\rm d\sim 8''$ of this galaxy. This is possible since virtually
all the bright $ ^{12}$CO J=1--0 emission lies within that region, and
hence the $  ^{12}$CO   J=3--2 emission  with  its higher   excitation
requirements will have, at  most, a similar  size.  Hence the observed
main-beam brightness   of the J=3--2 transition    can be expressed as
follows

\begin{equation}
\rm T_{\rm mb} = \left( \rm 1-e^{\rm -x^2} \right)\ T^{(\rm w)}_{\rm b }
\left(1+\frac{e^{\rm -x^2}}{1-e^{\rm -x^2}} \ \rm C \right)
\end{equation}

\noindent
where $\rm T^{(\rm w)} _{\rm  b}$ is the brightness temperature of the
warm gas  in the nucleus,  $\rm x=\sqrt{ln2}\ d/\Theta_{\rm  HPBW}$ (a
disk source  assumed), and $\rm C  = T^{(\rm c)}  _{\rm r}/T^{(\rm w)}
_{\rm r}$  is the  ratio of radiation  temperatures for J=3--2  of the
cold   and  warm   gas  phase.   The  former   dominates   at  larger
galactocentric   distances    in   this   galaxy    (Papadopoulos   \&
Seaquist~1998), and fills the rest of the 14$''$ beam.

Substituting $\rm d=8''$ and $\Theta _{\rm HPBW}=14''$ we obtain $ \rm
T^{(\rm w)} _{\rm b} = 4.94\left(1+3.94 \rm C\right)^{-1} T_{\rm mb}$,
which we plot in Figure 3 (for $\rm C=0$) together with the $ ^{12}$CO
J=1--0  brightness temperature  averaged  over the  inner diameter  of
$\sim 8''$.

\placefigure{fig3}

The  excellent  agreement  between  the  two  spectral  line  profiles
 demonstrates  that  they  arise  from  the  same  region,  which  the
 high-resolution $ ^{12}$CO J=1--0  maps (e.g. Tacconi \& Genzel 1996)
 allow  us to identify  with the  inner $8''$.   The velocity-averaged
 $\rm  R_{32}=(3-2)/(1-0)$ ratio  for that  region is  $\rm R_{32}\sim
 1.1$  (C=0).  If  the typical  physical characteristics  of  the cold
 phase, namely  $\rm n(H_2)\approx 3\times  10^2$ cm$ ^{-3}$  and $\rm
 T_{\rm kin}\approx  10 $~K (Papadopoulos \&  Seaquist 1998), dominate
 beyond   the  central   $8''$  while   warm  and   dense   gas  ($\rm
 n(H_2)>10^3$~cm$ ^{-3}$,  $\rm T_{\rm kin}>20$  K) is present  in the
 central region,  then $\rm C\sim  0.030-0.045$.  This yields  a range
 $\rm  R_{32}\sim 0.95-1.0$,  which for  a gaussian  source brightness
 will be somewhat lower, namely $\sim 0.80-0.85$.

Such high $\rm R_{32}$ ratios are characteristic of an optically thick
and thermalised  J=3--2 transition at gas temperatures  of $\rm T_{\rm
kin}\ga 30$ K,  consistent with the deduced dust  temperature of $\sim
35$   K.    A   simple   Large  Velocity   Gradient   (LVG)   analysis
(e.g. Richardson 1985) yields a variety of conditions that can produce
$\rm  R_{32}\sim 0.8-1.1$, all  of them  characterized by  $\rm T_{\rm
kin}\ga 30$ K and with the most typical temperature being $\sim 50$ K.
This suggests that most of  the molecular ISM in the starburst nucleus
of  NGC 7469  is indeed  warm with  its dust  emission  dominating the
global mm/sub-mm/FIR spectrum (Figure 2).

\section{Discussion}

The  high $\rm  R_{32}$ ratio  and  gas temperature  inferred for  the
central region  of NGC 7469  are in sharp  contrast with the  low $\rm
R_{21}=(2-1)/(1-0)\sim  0.5$   and  $\rm  T_{\rm   kin}\approx  10$  K
characterizing the  global CO emission from its  disk (Papadopoulos \&
Seaquist 1998).   Such steep excitation gradients  in spiral galaxies,
particularly the ones with starburst nuclei, are expected and known to
exist in numerous cases (Knapp et al 1980; Wall \& Jaffe 1990; Wall et
al.  1991;  Eckart et al 1991; Harris  et al. 1991; Wild  et al. 1992;
Aalto et al. 1995). This is further corroborated by the new generation
of mm and  sub-mm bolometers that allow sensitive  imaging of the dust
emission out  to large galactocentric distances and  show large masses
of cold dust  ($\rm T_{\rm d}=10-15$~K) residing in  spiral disks away
from the  nucleus (e.g.  Neininger et  al.  1996; Dumke  et al.  1997;
Alton et al.  1998; Papadopoulos \& Seaquist~1999a).

It is often argued that such a cold  dust component is responsible for
a  systematic underestimate of dust mass  when only FIR data are used.
This would  naturally  lead to the   large  $\rm M(H_2)/M_{\rm  d}\sim
500-600$ ratios ($\rm M(H_2+HI)\sim  1000$ when HI is included)  found
for spiral galaxies in contrast to $\sim 100-150$  found for the Milky
Way (e.g.  Devereux  \&  Young 1990  and references  therein).  On the
other  hand  extragalactic $\rm  M(H_2)$  estimates are  thought to be
accurate to within a factor of $\sim 2$ (Young  \& Scoville 1991), and
particularly for  spirals to  within $\pm   30\%$  (Devereux \&  Young
1990).  However recent sensitive sub-mm imaging at 450 $\mu $m and 850
$\mu $m of IR-bright spirals like NGC 891  (Alton et al. 1998) and NGC
1068 (Papadopoulos \& Seaquist  1999a) reveals that  the warm and cold
dust are spatially well separated,  with the former being concentrated
mainly in  the  inner $\la  2$ kpc   where significant  star formation
occurs and the latter residing at larger galactocentric radii.

The case  of NGC 1068 is  particularly relevant since  its host galaxy
properties are remarkably  similar to NGC 7469  (Wilson et al.  1991),
with both galaxies harboring  a  central starburst and having  similar
FIR, $ ^{12}$CO J=1--0 and HI luminosities.  The relative proximity of
NGC 1068  allowed  extensive multi-line CO and  sub-mm  imaging of its
inner  2.5~kpc (Papadopoulos  \&  Seaquist 1999a, 1999b)  which showed
that the bulk  of the gas  and dust in that region  is  warm with $\rm
M(H_2)/M_{\rm d}\sim 330$.  The latter is essentially identical to the
total gas-to-dust  ratio since  most  of the  gas in the circumnuclear
starburst of NGC 1068 is   molecular.  This  ratio is actually   $\sim
30-40\%$  higher still  if the  bright  $ ^{12}$CO J=3--2 emission  is
subtracted from the 850  $\mu $m flux used  to estimate the dust mass.
In the regions beyond the starburst nucleus  of NGC 1068 emission from
cold  dust ($\rm T_{\rm  d}\sim  10-15$~K) dominates the sub-mm bands,
while  low-brightness $ ^{12}$CO J=1--0  and bright HI emission reveal
the diffuse H$_2$ and HI that make  up the gas reservoir (Papadopoulos
\&  Seaquist 1999a).  Interestingly    the total gas-to-dust  ratio in
those regions is $\sim 70-150$, close to the Galactic value.

The larger  distance of NGC  7469 precludes a similar  detailed study,
however the  high-resolution $  ^{12}$CO J=1--0 and  450 $\mu  $m maps
(Figure  1) together  with  the lower  limit  of $\sim  35$~K for  the
dust/gas temperature in its central 8$''$ allow for a firm lower limit
on   the  gas-to-dust  ratio   in  that   region.   We   measure  $\rm
S_{450}(r\leq 4'')= (0.80\pm 0.08)$~Jy, probably a slight overestimate
owing to the larger beam area  at this wavelength coupling to a region
somewhat larger than  the inner 8$''$, beyond which  only faint sub-mm
emission  from  cold  dust  is expected.   The  velocity-integrated  $
^{12}$CO  J=1--0  line  flux  is $\rm  S_{\rm  CO}(r\leq  4'')=(175\pm
25)$~Jy~km~s$ ^{-1}$, estimated from the original high resolution map.
Assuming that H$_2$ dominates the gas phase in the starburst region, a
Galactic $\rm  X_{\rm CO}$, and $\rm  T_{\rm d}=35$ K,  yields a total
gas-to-dust ratio of~$\sim 500$.

Hence for  both NGC 7469  and NGC 1068  it can be  convincingly argued
that  the   high  gas-to-dust  ratios  found   for  their  IR-luminous
starbursting  central   regions  are  not  due  to   the  presence  of
significant  amounts  of cold  dust  but  to  an overestimate  of  the
molecular gas mass in such environments.  In the case of NGC 7469 this
conclusion is  further supported by  a dynamical study of  its central
region  concluding that $\rm  X_{\rm CO}$  is $\sim  1/5$  of the
Galactic value (Genzel et al. 1995).  This would bring the gas-to-dust
ratio  estimated for that  region in  good accord  with the  Milky Way
value with no need for an unaccounted mass of cold~dust.

\subsection{The systematic overestimate of $\rm M(H_2)$ in starburst
environments}

There  has been  mounting evidence  that  the Galactic  value of  $\rm
X_{\rm  CO}$  overestimates the  molecular  gas  mass  in the  intense
starburst  environments of luminous  IR galaxies  ($\rm L_{\rm  FIR} >
10^{11}\ L_{\odot  }$ by a  factor of $\sim  5$ (Solomon et  al. 1997;
Downes  \&  Solomon  1998).  This  seems  to  be  due to  a  two-phase
differentiation that the molecular  gas undergoes in such environments
(Aalto et al. 1995; Downes \& Solomon 1998).  The phase that dominates
the   $  ^{12}$CO  emission   is  diffuse,   warm  and   possibly  non
self-gravitating,   the  latter   being  the   main  reason   for  the
overestimate of  molecular gas when  a Galactic value for  $\rm X_{\rm
CO}$ is used.

 Indeed  a standard  expression for  this conversion  factor  is (e.g.
Bryant \& Scoville 1996)

\begin{equation}
\rm X_{\rm CO}=2.1\  \frac{n^{1/2}}{\rm  T_{\rm
b}}\ Q\ M_{\odot }\ (K\ km\ s^{-1})^{-1}
\end{equation}

\noindent
where n  (cm$ ^{-3}$) is the  average gas density and  $\rm T_{\rm b}$
(K) is the average brightness  temperature of $ ^{12}$CO J=1--0 of the
molecular cloud  ensemble.  The factor $\rm Q=  \delta V_{vir} /\delta
V$ accounts for the non-virial linewidth of the ``average'' cloud, for
self gravitating clouds it is $\rm Q=1$.

For typical  conditions in the  Galaxy, namely $\rm Q\approx  1$, $\rm
n\approx  300$ cm$  ^{-3}$  and $\rm  T_{\rm  kin}\approx 15$~K  ($\rm
T_{\rm  b}  \approx 8$  K),  one  obtains  $\rm X_{\rm  CO}\approx  5\
M_{\odot }\ (K\ km\ s^{-1})^{-1}$, the typical Galactic value.  In the
diffuse molecular phase that is present in starburst nuclei, it can be
$\rm Q<1$  (e.g.  Solomon  et al. 1997),  thus a Galactic  $\rm X_{\rm
CO}$ will overestimate the molecular gas mass present.  Moreover, even
in the case of self-gravitating clouds, a diffuse ($\rm n\approx 10^3$
cm$ ^{-3}$) but warm phase ($\rm T_{\rm kin}\sim 60$ K) dominating the
$ ^{12}$CO  J=1--0 emission will  have $\rm T_{\rm b}=30-40$  K ($\tau
_{10}\sim 2-4$) and thus yield a $\rm X_{\rm CO}$ factor that is $\sim
2-3$ times smaller than the Galactic value.

If  such a gas  phase is  characteristic of  the starburst  regions in
 luminous IR  galaxies, then a {\it systematic}  overestimate of H$_2$
 mass rather than an underestimate of dust mass is responsible for the
 high  $\rm M(H_2)/M_{\rm  d}$  ratios found  for  them.  Because  the
 effect is  systematic it  can produce both  a high  total gas-to-dust
 ratio and  a small dispersion  around its mean. Hence  earlier claims
 that such small dispersion  ($\sim \pm 30\%$) observed in IR-luminous
 spirals  (Devereux  \& Young  1990)  argues  in  favor of  a  similar
 accuracy in molecular  gas   mass  estimates   are  not   supported  by
 this~picture.

\subsection{The cold gas and dust in NGC 7469}

The global excitation of CO in NGC 7469 seems  to be dominated by cold
and  diffuse  molecular  gas  out   to a  scale  of  $\rm  d\la 20''$
(Papadopoulos \&  Seaquist 1998).  A  gaussian fit  of the 850~$\mu $m
emission (Figure 1) yields a source size of $\sim 17.2''\times 18.5''$
(FWHM), clearly larger than the  beam at this wavelength.  Considering
the geometric mean of  $\Theta _{\rm s} \sim 18''$  to be the observed
diameter of the source (assumed to be a disk), and for a gaussian beam
with $\Theta_{\rm HPBW}=15''$,  we obtain an intrinsic source diameter
of $\rm d=\left [2\ (ln 2) ^{-1}\ (\Theta ^2  _{\rm s}-\Theta ^2 _{\rm
HPBW})\right]^{1/2}\sim 17''$.  This is $\sim 2$ times larger than the
size of  the bright $ ^{12}$CO  J=1--0 emission  and comparable to the
source size seen at 450 $\mu $m (Figure~1).

In  NGC  1068  the warm  gas/dust  resides  in  the inner  $\rm  d\sim
(2.7-3.4)$~kpc, which is  comparable to the size of  a similar region in
NGC~7469, both regions  containing an embedded starburst. Furthermore,
given  the  similarities of  the  disk  properties  between these  two
galaxies, it  is reasonable to  identify the extended faint  sub-mm, $
^{12}$CO J=1--0 emission in NGC~7469 with emission from cold dust/gas.
Indeed simply scaling the total size of the faint sub-mm emission from
cold  dust in  NGC 1068  ($\sim 6$  kpc) to  the distance  of NGC~7469
yields  an angular  size  of  $\sim 20''$,  closely  matching the  one
observed for the latter.

It is intriguing that faint $ ^{12}$CO J=1--0 emission is detected out
to significantly  larger radii  than the  inner radius  of $\rm r=4''$
(Figure 1).   Assuming that, as in  NGC  1068, this emission ``marks''
cold   and diffuse  H$_2$   coexisting with  cold  dust  ($\rm  T_{\rm
d}=10-15$    K)  and   the   bulk   of  HI,     yields  a ratio   $\rm
[M(H_2+HI)]/M_{\rm d} \sim   50-160$,  close to the Milky Way value.

\subsection{Warm versus cold gas/dust:  sub-mm imaging and CO
spectroscopy}

It is striking that in  the spatially integrated dust emission of this
galaxy the cold dust is  inconspicuous even after the inclusion of the
sub-mm data. On the other hand its molecular gas counterpart dominates
the  global  excitation  characteristics  as  revealed by  the  low  $
^{12}$CO  (J=2--1)/(J=1--0)  line  ratio (Papadopoulos  \& Seaquist 1998).

This could simply be due to  the fact that spectral line luminosity is
sensitive  to the local  gas density  as well  as its  temperature and
total mass but only the  latter two determine the continuum luminosity
from  dust  (for a  fixed  gas-to-dust  ratio).   In other  words  two
molecular clouds with the same mass and gas/dust temperature will emit
the same dust continuum (optically thin case) but their CO line fluxes
can differ drastically if their average H$_2$ densities are different.
This will occur mainly in the density regime of sub-thermal excitation
of the observed CO line(s).

In  NGC  7469 the  relative  amounts  of warm  and  cold  dust can  be
estimated from the relation

\begin{equation}
\rm m= \frac{M^{(c)} _{\rm d}}{M^{(w)} _{\rm d}} = \left[ \frac{\rm e^{\rm 32/T_{\rm c}}-1}{\rm e^{\rm 32/T_{\rm w}}-1}\right] \frac{\rm S^{(\rm c)} _{450}}{
\rm S^{(\rm w)} _{450}},
\end{equation}

\noindent
where (w) and (c) denote  the quantities corresponding to the warm and
cold dust respectively. It is  $\rm S^{(\rm w)} _{450} = S_{450}(r\leq
4'') =  0.8$ Jy  and $\rm S^{(\rm  c)} _{450} =  S_{450}(4''\leq r\leq
11'') =0.5$  Jy. Hence  for $\rm  T_{\rm w}\ga 35$  K and  $\rm T_{\rm
c}=10-15$ K we obtain $\rm m \ga 3-10$.

Even for $\rm  m\sim 10$, the spatially averaged  sub-mm emission will
not  necessarily reveal the  presence of  the cold  component. Indeed,
since the ratio $\rm r= S_{450}/S_{850}$ is temperature-sensitive when
$\rm  T_{\rm d}\la 30$  K, it  is expected  to differ  if a  cold dust
component  is  present along  with  the  warm  one.  Assuming  spatial
averaging of the emission from  the two dust components the observed r
can be written as follows

\begin{equation}
\rm r= r_{\rm w} \left[ \frac{\rm 1+m\ f_{450} (T_w, T_c)}{\rm 1 + m\ f_{850}(T_w, T_c)}\right],
\end{equation}

\noindent
where $\rm r_{\rm w}$ is the ratio corresponding to warm dust alone and, 

\begin{equation}
\rm f_{\lambda}(T_w, T_c) = \frac{\rm e^{\rm T_{\lambda }/T_{\rm w}}-1}{
\rm e^{\rm T_{\lambda }/T_{\rm c}}-1}, \ \ 
T_{\lambda } = \frac{\rm\ h\ c}{\lambda \ k}.
\end{equation}

For $\rm  T_{\rm w}=40$  K, $\rm T_{\rm  c}=10$ K  and $\rm m=  10$ we
obtain  $\rm  r  \approx   0.70\times  r_{\rm  w}$,  which  is  barely
discernible from $\rm  r_{\rm w}$ given that the value  of r carries a
$\sim 20\%$ uncertainty due  to the calibration uncertainties of SCUBA
at 450 $\mu $m and 850 $\mu $m.

On the  other hand  besides temperature the  spectral line  ratios are
also  sensitive  to the  local  gas density.   Thus  a  cold {\it  and
sub-thermally excited}  molecular gas phase may  dominate the observed
global  line ratio  even in  the presence  of a  warm  and thermalized
phase.  For example, the  global $\rm R_{21}=(2-1)/(1-0)$ ratio can be
expressed as

\begin{equation}
\rm R_{21} = R^{(w)} _{21}\left[ \frac{\rm 1+m\ x\ (\rm R^{\rm (c)} _{21}/R^{\rm (w)} _{21})}{\rm 1 + m\ x}\right],
\end{equation} 

\noindent
where  $\rm x =  X^{(\rm w)}_{\rm  CO}/X^{(\rm c)}  _{\rm CO}$  is the
ratio  of the  CO(1--0)-to-H$_2$ conversion  factors for  the  two gas
phases.  Assuming  $\rm x\sim 1/5$,  $\rm m\sim 10$ and,  $\rm R^{(\rm
w)} _{21}  \sim 1$, we obtain  $\rm R_{21}\sim 1/3\  (1+2\ R^{(\rm c)}
_{21})$.

In NGC  7469 it is  $\rm R_{21}\approx 0.5$ (Papadopoulos  \& Seaquist
1998),  hence   the  cold-phase  ratio   will  be  $\rm   R^{\rm  (c)}
_{21}\approx  0.25$,  and  for  $\rm  T_{\rm kin}=10$  K,  the  latter
corresponds to densities of  $\rm n(H_2)\approx 10^2$ cm$ ^{-3}$.  For
the same  temperature but thermalized  and optically thick  $ ^{12}$CO
J=2--1  it  is  $\rm  R^{\rm  (c)} _{21}\approx  0.8$,  yielding  $\rm
R_{21}\approx 0.85$ which is significantly larger than observed.

The  aforementioned simple analysis  demonstrates the  significance of
combining  sub-mm  and  CO   observations  in  evaluating  the  physical
conditions  and the mass  of molecular  gas.  The  sub-mm measurements
offer an independent means of estimating gas mass under the assumption
of a  canonical gas-to-dust ratio  of $\sim 100$.  For H$_2$-dominated
gas  this mass can  then be  compared to  the one  deduced from  the $
^{12}$CO J=1--0 luminosity and  the $\rm X_{\rm CO}$ conversion factor
and hence the influence of the physical conditions on $\rm X_{\rm CO}$
can be assessed.   On the other hand, if a  cold and diffuse molecular
gas phase  is present along with  a warm one but  only global averages
are available, CO line ratios can be more sensitive to the presence of
the cold ISM phase than sub-mm intensity ratios.

\section{Conclusions}

We presented sensitive 850 $\mu $m,  450 $\mu $m maps of the Seyfert 1
galaxy  NGC 7469  and a  measurement of  the $  ^{12}$CO  J=3--2 line
towards its starbursting central region. Our main conclusions can be
summarized as follows

1. The FIR/sub-mm/mm  spectrum of this  source is dominated by  a warm
dust component  with a  temperature of $\sim  35$ K and  a global
molecular gas-to-dust ratio of $\sim 600$, both typical for IR-luminous 
spirals.  Including the atomic gas mass  yields a total gas-to-dust ratio
of $\sim 800$.

2. The warm  dust and gas   lies in the inner   8$''$ ($\sim $2.5 kpc)
 where a  starburst is embedded.   In this region  we find no evidence
 for  a significant  mass of  cold dust, yet   the ratio of the mainly
 molecular gas to the dust mass is $\sim 500$, still five times larger
 than the  Galactic value.  We argue  that this  is   the result  of a
 systematic overestimate  of $\rm  H_2$ mass  by a  factor of $\sim 5$
 when a Galactic value for $\rm X_{\rm  CO}=M(H_2)/L_{\rm CO}$ is used
 in starburst environments.

3. On larger  scales (radius  $\ga 1.2$  kpc) the ISM  in NGC  7469 is
dominated by cold, sub-thermally excited  gas where the faint 450 $\mu
$m and $ ^{12}$CO J=1--0 emission originate. The gas-to-dust ratio for
this   phase,  is   $\sim   50-150$;  in   better   accord  with   the
Milky~Way~value.

\subsection{Acknowledgments}

We would like to thank Linda  Tacconi for providing us with data prior
to publication, Henry Matthews for  obtaining the SCUBA maps, and Remo
Tilanus    for    conducting    the    CO    observations    on    our
behalf.  P. P.  Papadopoulos is  supported by  the ``Surveys  with the
Infrared Space Observatory'' network set up by the European Commission
under contract FMRX-CT96-0086 of its TMR programme.

\newpage

\appendix

\section{Spectral line contribution to mm/sub-mm bands}

In mm and sub-mm continuum  observations one must correct for non-dust
 emission  contributions. The  most significant  comes  from molecular
 spectral lines  and can  reach up  to $\sim 60\%$  of the  total flux
 observed with  a typical bolometer bandwidth (Gordon  1995).  In warm
 cores of Orion  such line contributions are found to  be of the order
 of $\sim  30\%$ in 850~$\mu  $m and 450  $\mu $m bands  (Johnstone \&
 Bally  1999 and  references therein).   However in  the extragalactic
 domain  a typical  beam of  a  mm/sub-mm telescope  samples areas  of
 several hundred parsecs, and the contributions from warm cores ($\leq
 1$  pc)  in  star  forming  regions are  negligible  because  of  the
 resulting spatial dilution.  Over  such scales only the emission from
 the  three  lowest  rotational  transitions  of  $  ^{12}$CO  can  be
 ubiquitous and bright, since  these transitions can be easily excited
 in  the general  conditions prevailing  in  the ISM  in quiescent  or
 starburst environments.

In the  case of NGC 7469  we expect that all  the significant spectral
line emission from the $ ^{12}$CO  J=2--1 (1.3 mm band) and $ ^{12}$CO
J=3--2 (850  $\mu $m band)  lines arises from  the region of  bright $
^{12}$CO  J=1--0 emission  in the  central 8$''$  where  the starburst
resides. In the  case of the 1.3 mm band  only line-free channels were
used to  produce the  continuum map (Tacconi,  private communication).
The  850  $\mu $m  band  includes the  $  ^{12}$CO  J=3--2 line  whose
contribution  can  be estimated  from  our  observation  of this  line
towards the nucleus, namely

\begin{equation}
\rm S_{850 } ^{\rm (dust)} = S_{850 }  - \frac{2\ k \ \nu  _{\circ }
^3}{c^3\    \Delta   \nu   _{B}}\left(\frac{\rm    x^2}{\rm   1-e^{\rm
-x^2}}\right) I_{\rm mb}\ \Omega _{\rm mb}
\end{equation}

\noindent
where $\rm  I_{\rm  mb}$ (K  km s$  ^{-1}$) is the velocity-integrated
main-beam brightness of $ ^{12}$CO  J=3--2, $\nu _{\circ } =  345.796$
GHz, and   $\Omega  _{\rm mb}$  is  the  gaussian  beam area   at this
frequency.  The term in  the parenthesis corrects for  the beam-source
geometrical coupling assuming a disk source  of diameter d, where $\rm
x=\sqrt{ln2}\ d/\Theta _{\rm HPBW}$, and $\Delta  \nu _{\rm B}$ is the
bolometer bandwidth.

For $ ^{12}$CO J=3---2 in  NGC 7469 we measured $\rm I_{\rm mb}=(77\pm
8)$ K  km s$ ^{-1}$ (Figure  3), hence for  $\Theta _{\rm HPBW}=14''$,
$\Delta  \nu _{\rm B}=  30$ GHz  (Holland, private  communication) and
$\rm d=8''$ we  obtain a line contribution of $\Delta  \rm S= 63$ mJy,
which amounts to $\sim  40\%$ of the total flux at 850  $\mu $m. The $
^{13}$CO J=3--2 is expected to have $\la 0.1$ of the $ ^{12}$CO J=3--2
flux, and hence contribute $\la 4\%$ of the total flux in 850 $\mu $m.

 In  the case  of the  450 $\mu  $m  band the  only CO  line that  may
contribute  to the  observed  flux  is $  ^{13}$CO  J=6--5 at  661.067
GHz. However we find that  this line, with its high excitation density
($\rm n>10^5$  cm$ ^{-3}$), will  not contribute significantly  to the
450  $\mu $m band  (i.e $\la  2\%$) over  the spatial  scales relevant
here, even in a starburst environment.

{}

\newpage

\figcaption{Top: the 850 $\mu $m map with a FWHM= 15$''$ beam shown at
the lower left, the  contours are $(-3, 3, 6,  9, 12, 15, 18,  21, 24,
27,   30)\times  \sigma _{850}$, with  $\sigma   _{850} = 5$ mJy beam$
^{-1}$.  \\ Bottom:  the 450 $\mu  $m  map (contours)  overlaid to the
integrated $   ^{12}$CO  J=1--0  emission (grey  scale)   at  a common
resolution of 9$''$ (FWHM beam shown at the bottom left). The contours
are $(-4, 4, 6, 8, 10, 14, 16, 18)\times  \sigma_{450 }$, with $\sigma
_{450 }= 40$ mJy beam$ ^{-1}$ and the  grey scale is I=6--200 Jy beam$
^{-1}$  km s$  ^{-1}$ ($\sigma (\rm  I)=  1\rm \  Jy\ beam ^{-1}\  km\
s^{-1}$) \\ The map center is at  RA: 23$ ^{\rm  h}$ 03$ ^{\rm m}$ 15$
^{\rm s}$.6, Dec: +08$^{\circ }$ 52$'$ 26$''$ (J2000).
\label{fig1}}

\figcaption{A $\chi ^2$-fitted  SED for the  mm/sub-mm/FIR data of NGC
7469 (Table 1).  An optically thin isothermal  dust reservoir has been
assumed with  an emissivity  law of $\alpha=2$   and emissivity at 200
$\mu $m (1196 GHz)  of  $\rm k _{\circ }=  10  $ cm$ ^{2}$ gr$  ^{-1}$
(Hildebrand 1983).  The flux at 25 $\mu $m is not used in the fit.
\label{fig2}}

\figcaption{The $  ^{12}$CO J=1--0 and  J=3--2 brightness temperatures
for  the  inner  $8''$  of  NGC~7469 (see  text).   The  thermal  rms
uncertainty across  the band  is $\rm \delta T_{\rm  rms} (1-0)=30 $  mK, and
$\rm \delta T_{\rm rms} (3-2)=40$ mK.
\label{fig3}}

\newpage

\centerline{\large Table 1}
\vspace*{0.5cm}
\centerline{\large NGC 7469: The mm, sub-mm and FIR data}
\begin{center}
\begin{tabular}{ c c c }\hline\hline

Wavelength & Flux  & Reference \\ \hline

1.3 mm       &  $19\pm 5$ mJy     &  Tacconi et al. 1999 $ ^{\rm a}$ \\
850 $\mu $m  &  $91 \pm 13$ mJy   & this work $ ^{\rm b}$ \\
450 $\mu $m  &  $1.30\pm0.14$ Jy & this work $ ^{\rm b}$ \\
100 $\mu $m  &  $34.9\pm 0.6$ Jy  & Soifer et al. 1989 \\
60  $\mu $m  &  $27.7\pm 0.04$ Jy &   `` `` \\ 
25  $\mu $m  &  $5.84\pm 0.05$ Jy &   `` `` \\ \hline
\end{tabular}
\end{center}

\noindent
$ ^{\rm a}$ Derived from the 1.3 mm continuum map for a radius of $\rm
r=4''$ that contains all the mm emission and corrected
for non-thermal emission using the data from Wilson et al. 1991.

\noindent
$ ^{\rm b}$ Fluxes are estimated for a radius of $\rm r= 11''$ and the
 appropriate  error-beam   corrections  (Sandell  1997)   are  $\rm  f
 _{850}=0.80$, $\rm  f _{450}=1.05-1.10$, derived from  Uranus and CRL
 618  maps.  The  850  $\mu  $m  flux  has  also  been  corrected  for
 contribution  from the  $ ^{12}$CO  J=3--2 line  (see  Appendix). The
 errors  reported include  (besides the  thermal rms  errors)  a $\sim
 $10\%  (850 $\mu  $m) and  a  $\sim $15\%  (450~$\mu $m)  calibration
 uncertainty.

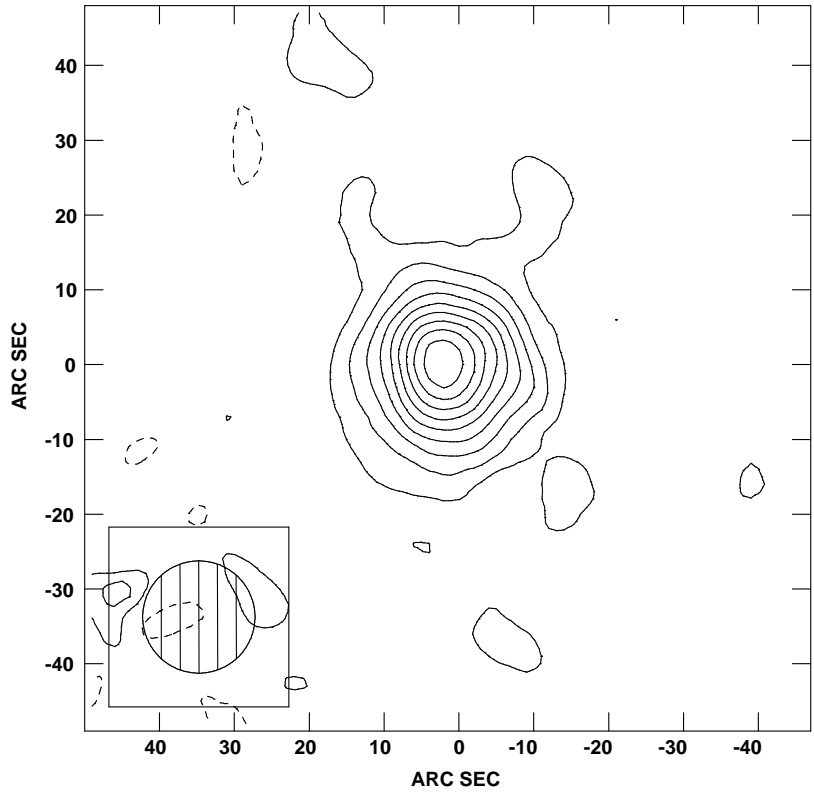
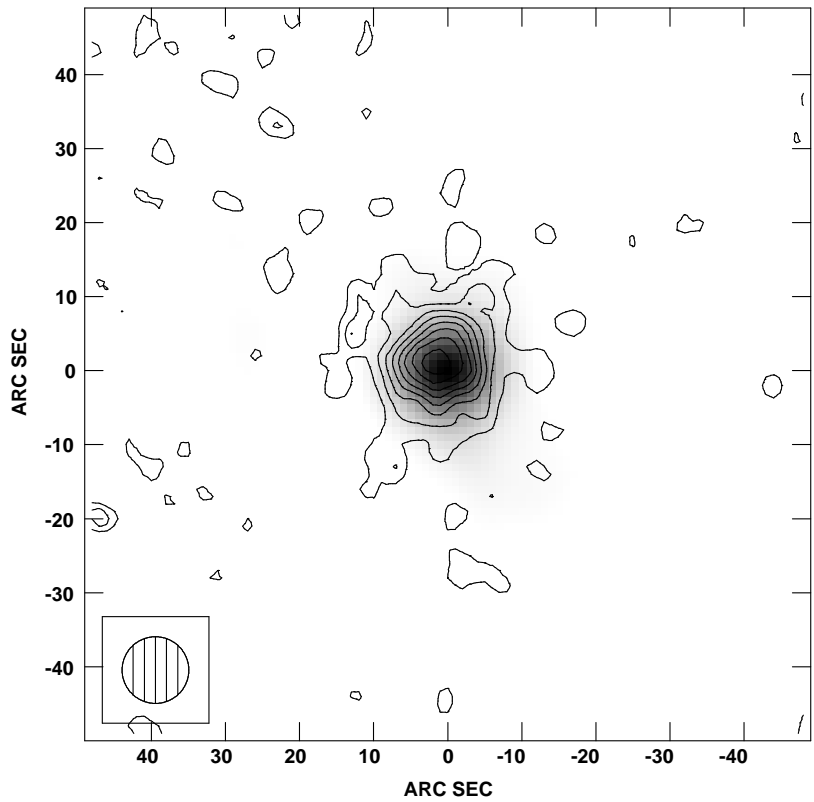
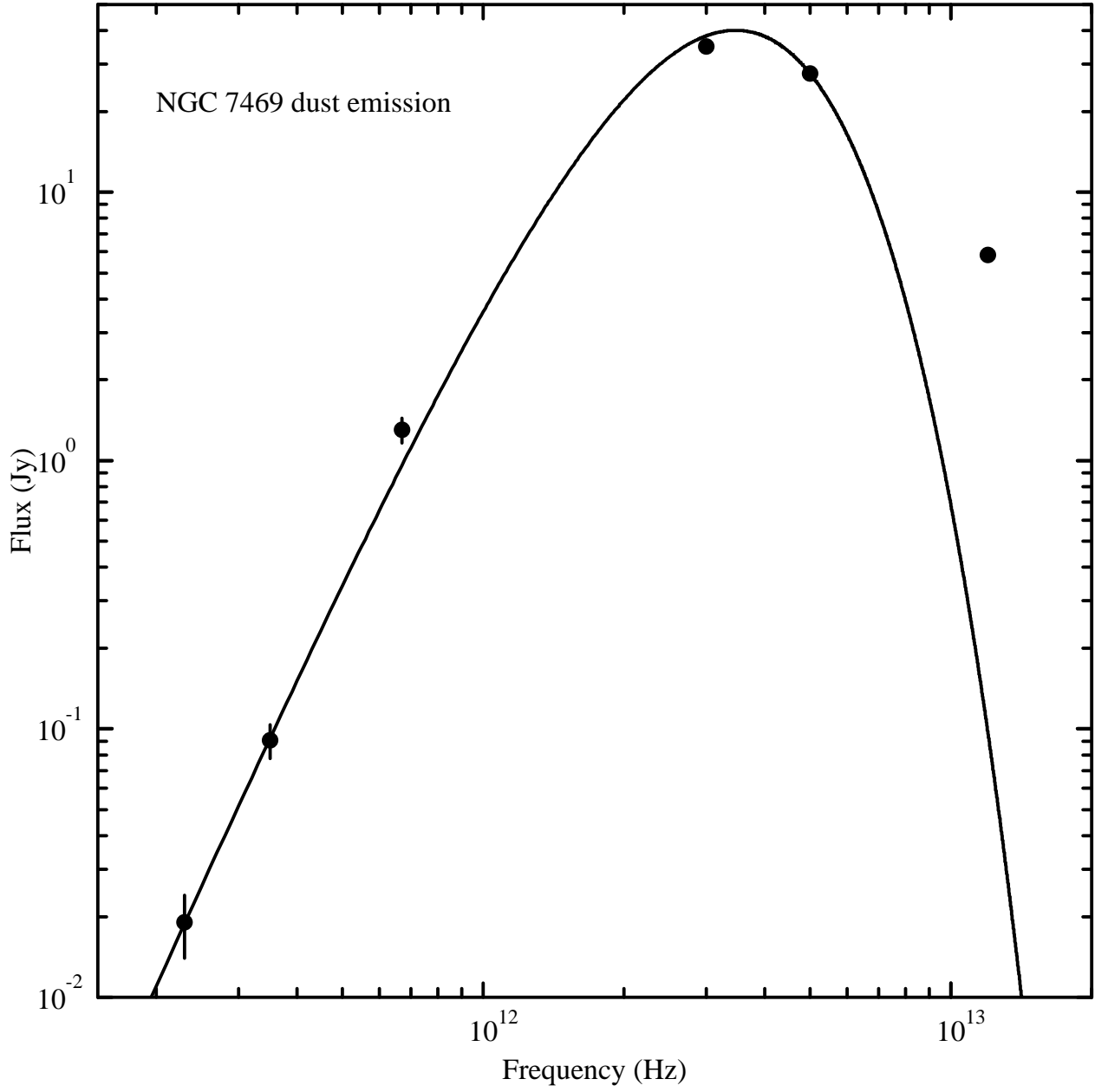
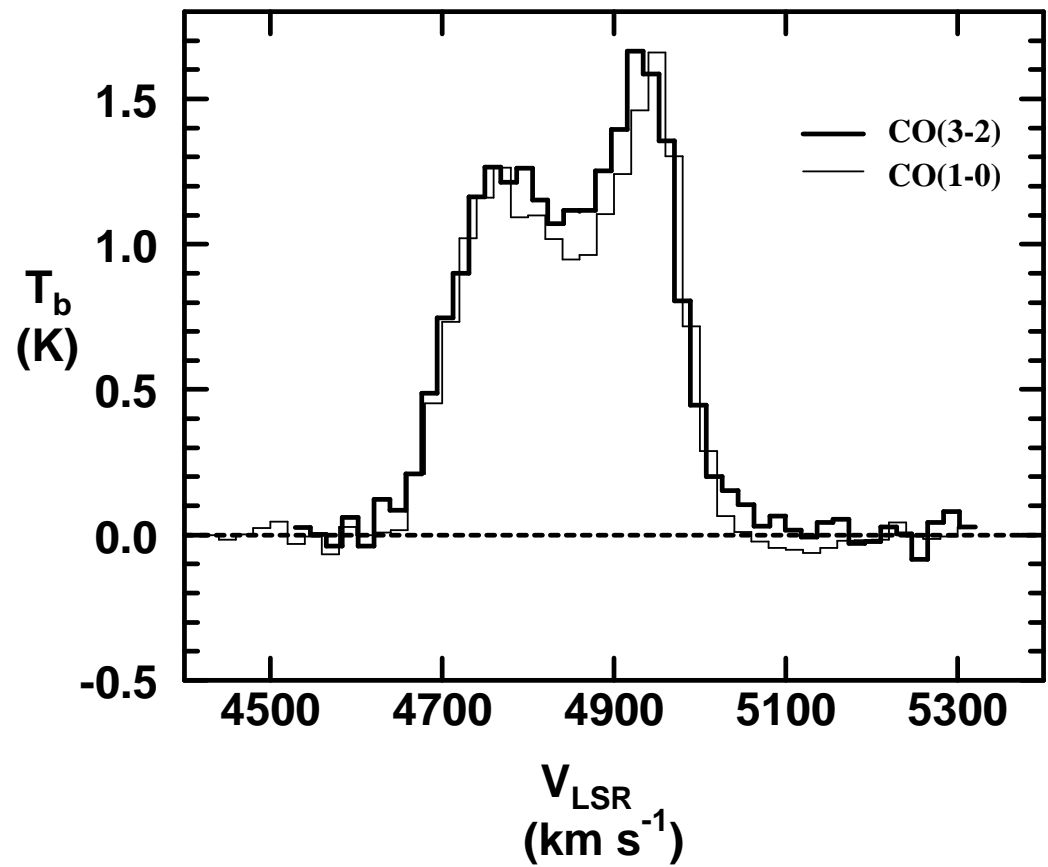
\end{document}